\documentclass[5p,number]{elsarticle}
\usepackage{hyperref}
\usepackage[T1]{fontenc}
\usepackage[utf8]{inputenc}
\usepackage{textcomp} 
\usepackage{lmodern} 
\usepackage{amsmath}
\usepackage{xcolor}
\usepackage{amsfonts}
\usepackage{breakurl} 
\usepackage[switch]{lineno} 
\graphicspath{{./figures/}}
\biboptions{sort&compress}
\usepackage{subcaption} 
\usepackage{booktabs}
\graphicspath{{./figures/}}
\usepackage{dcolumn}
\usepackage{multirow}
\usepackage{adjustbox}
\usepackage{bm}
\modulolinenumbers[5]
\usepackage{textgreek}

\hypersetup{
    breaklinks = true,
    colorlinks = true,
    pdftitle = {}, 
    pdfauthor = {},
    pdfkeywords = {},
    linkcolor = blue,
    citecolor = blue,
    filecolor = black,
    urlcolor = magenta
}

\makeatletter 
\def\@author#1{\g@addto@macro\elsauthors{\normalsize%
    \def\baselinestretch{1}%
    \upshape\authorsep#1\unskip\textsuperscript{%
      \ifx\@fnmark\@empty\else\unskip\sep\@fnmark\let\sep=,\fi
      \ifx\@corref\@empty\else\unskip\sep\@corref\let\sep=,\fi
      }%
    \def\authorsep{\unskip,\space}%
    \global\let\@fnmark\@empty
    \global\let\@corref\@empty  
    \global\let\sep\@empty}%
    \@eadauthor={#1}
}
\makeatother

\begin{document}
\begin{sloppypar} 

\title{Magnetic properties and structural phase transition in ultrathin fcc Fe(111) and bcc Fe(111) films: first-principles study}

\author{Jakub Meixner$^{1,2}$\corref{cor1}}
\cortext[cor1]{Corresponding author} 
\ead{meixner@ifmpan.poznan.pl}

\author{Justyna Rychły-Gruszecka$^{1}$}
\author{Mirosław Werwiński$^{1}$}

\address{$^1$Institute of Molecular Physics, Polish Academy of Sciences,\\  M. Smoluchowskiego 17, 60-179 Poznań, Poland\\ $^{2}$Poznan University of Technology, Faculty of Materials Engineering and Technical Physics,\\ Pl. Marii Skłodowskiej-Curie 5, 60-965 Poznań, Poland}

\begin{abstract}
	The aim of this work is to investigate the structural and magnetic characteristics of Fe thin films with a triangular (hexagonal) lattice surfaces (fcc (111) and bcc (111)). 
	The properties of these structures have been calculated using density functional theory (DFT) implemented in the full-potential local-orbital(FPLO) code. 
	The results indicate a structural phase transition from fcc to bcc structure when the film thickness exceeds 23 Fe atomic monolayers.
	The considered fcc films prefer the low-spin ferromagnetic state with an average magnetic moment of about 1.0 $\mu_{B}$ per atom.
	This moment decreases with increasing film thickness until the critical thickness, where, after the structural transition to the bcc phase, it reaches a value close to that of bulk bcc Fe. 
	Moreover, the values of the magnetic anisotropy energy are positive (perpendicular magnetic anisotropy) for the entire thickness range of films with fcc structure (in ferromagnetic low-spin state) and systematically decrease with increasing film thickness.
	The presented computational results explain the experimentally observed structural transition and may help to select appropriate substrates with suitable lattice parameters for the deposition of ultrathin Fe(111) films.

\end{abstract}

\date{\today}

\maketitle

\section{Introduction}

\begin{figure}
\centering
\includegraphics[clip, width=0.92\columnwidth]{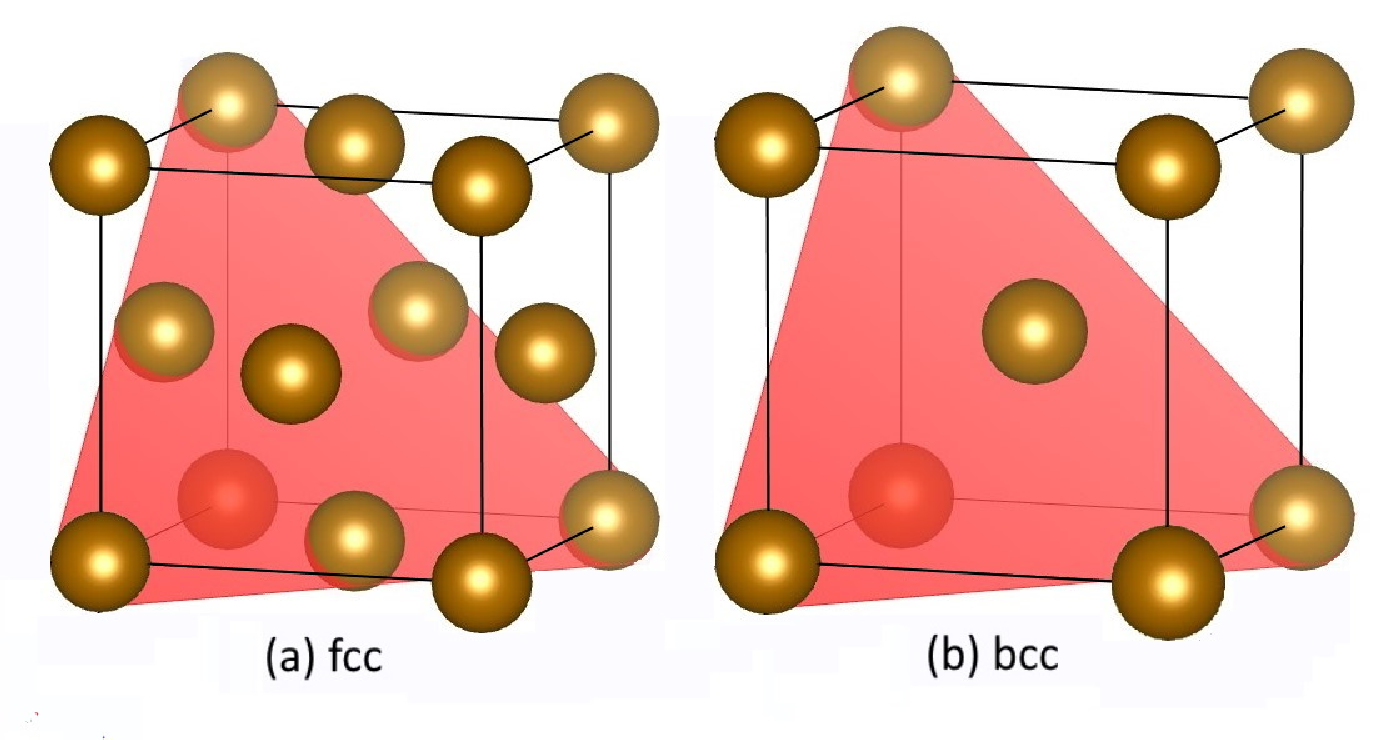}
\caption{\label{obraz0} Unit cells of considered Fe structures with marked (111) planes. (a)\,Face centered cubic (fcc) unit cell, $a_{bulk.fcc} = 3.65$~\AA. (b)\,Body centered cubic (bcc) unit cell, $a_{bulk.bcc} = 2.82$~\AA.}
\end{figure}

\begin{figure*}
\centering
\includegraphics[clip, width=0.92\textwidth]{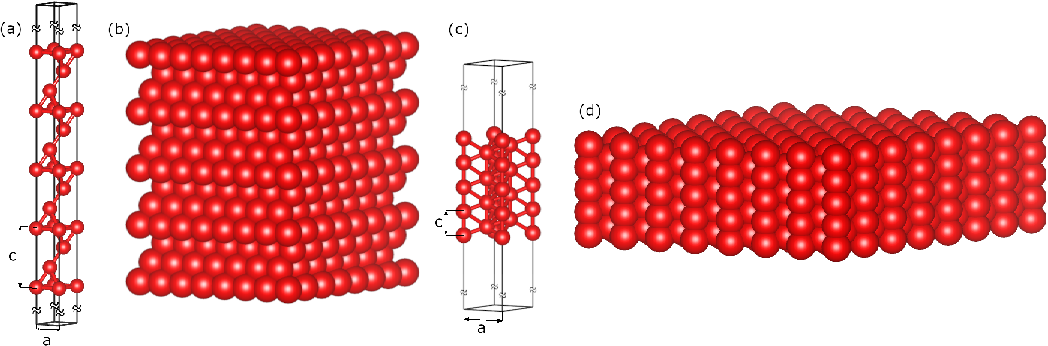}
\caption{\label{obraz1}Crystal structure models of 13 atomic monolayer thick Fe films with triangular lattice surfaces, space group \textit{P}-3\textit{m}1, as obtained after geometry optimization. 
(a)\,Unit cell of film with $a \approx 2.6$~\AA{} (fcc (111)).
(b)\,Multiplication of the fcc Fe(111) unit cell.
(c)\,Unit cell of film with $a \approx 4.0$~\AA{} (bcc (111)). 
(d)\,Multiplication of the bcc Fe(111) unit cell.}
\end{figure*}

	Iron is a transition metal important in many industrial and technological applications. 
	Under normal conditions, bulk Fe crystallizes in a body-centered cubic (bcc) lattice.
	The face-centered cubic (fcc) lattice of iron, see Fig.~\ref{obraz0}, is observed in ultrathin films and in bulk samples under certain pressure and thermal conditions (e.g. above 912$^{\circ}$C at atmospheric pressure) \cite{marynowska_structural_2016-1}. 
	The structural stability of Fe thin film is sensitive to the type of surface on which it is deposited \cite{hardrat_complex_2009-1}. 
	Ultrathin Fe films can be grown, among others, on substrates with square lattice (e.g., fcc (001) and bcc (001)) or hexagonal/triangular lattice (e.g., fcc (111), bcc (111), and hcp (0001)) surfaces \cite{kalki_evidence_1993-1, schmailzl_structure_1994-1, hoff_critical_1995-2, kolaczkiewicz_growth_1999-2, kronlein_magnetic_2018-2}.
	Fe(111) layers with a triangular lattice surface describe bcc or fcc structures, depending on the lattice parameter \textit{a}, see Fig.~\ref{obraz1}.
	
	Ultrathin Fe films growing on a substrate with a hexagonal-lattice surface, such as Au(111), exhibit an fcc structure, and an increase in Fe film thickness leads to a transformation to a bcc structure \cite{stroscio_microscopic_1992-1}.
	Depending on the type of surface on which the Fe film is grown, the critical film thickness at which the transition to the bcc phase is observed ranges from about 10 to 20 atomic monolayers \cite{kalki_evidence_1993-1, schmailzl_structure_1994-1, hoff_critical_1995-2,roldan_cuenya_observation_2001}.
	Since ultrathin Fe films are always deposited on a substrate, the common interpretation of the properties of such films also refers to the properties of the substrate.
	In particular, the fcc phase stabilization of ultrathin Fe(111) films is customarily interpreted as a substrate effect.

	The fcc structure of bulk Fe can exhibit several magnetic states such as spin spiral, ferromagnetic, or antiferromagnetic \cite{jiang_carbon_2003}. 
	First-principles calculations have shown that the ferromagnetic state in fcc Fe can take low-spin (high-spin) forms with magnetic moment 1.02 $\mu_B$ (2.57 $\mu_B$) and lattice parameter 3.49~\AA{} (3.64~\AA{}) \cite{jiang_carbon_2003}, among which the low-spin state is more stable.
	Experimentally determined value of the lattice parameter of the ferromagnetic fcc Fe phase is 3.65~\AA{} at 4 K \cite{acet_high-temperature_1994}.
	The magnetic low-spin state in Fe fcc thin films has also been observed experimentally \cite{shen_structural_1997, torija_frozen_2005}.
	
	Ultrathin Fe(111) films have been considered by theoreticians with the use of \textit{ab initio} calculations \cite{zhang_first_2011-1, shi_first-principles_2018-1, wu_structural_1993-1, zhuang_first-principles_2021-1, kruger_magnetic_2001}.
	Although Fe(111) films with hexagonal-lattice surfaces can occur in fcc \cite{zhang_first_2011-1, shi_first-principles_2018-1, kruger_magnetic_2001} and bcc \cite{wu_structural_1993-1, zhuang_first-principles_2021-1} structures, the authors of the computational works consider an arbitrarily chosen case.
	Wu and Freeman's 1993 paper on bcc (111) Fe layers \cite{wu_structural_1993-1} is a pioneering work considering films up to 13 atomic monolayers thick.
	Subsequent work has focused on small clusters (up to 10 atoms) of Fe \cite{bergman_magnetic_2007-1}, hexagonal Fe monolayer on substrate \cite{simon_spin-correlations_2014-2}, carbon-vapor reactions on the fcc-Fe(111) surface \cite{zhang_first_2011-1},  Fe interfaces with h-WC \cite{shi_first-principles_2018-1}, Al$_{2}$O$_{3}$ \cite{wang_first_2019-1}, TiB$_{2}$ \cite{wang_first_2021-2} and FeWB \cite{zhuang_first-principles_2021-1}, as well as the Fe-Fe interface: fcc Fe(111)/hcp Fe~(0001) \cite{lee_austenitee_2012}, among others.

	The aim of our study is to investigate the properties of the magnetic Fe film without substrate and overlayer, which are difficult to obtain experimentally.
	The present work focuses on the magnetic properties of ultrathin iron films with trigonal symmetry space group \textit{P}-3\textit{m}1. 
	We consider films with a triangular lattice surface that are known to form an fcc structure and are stable for thicknesses of a few monolayers. 
	In the limit of large thicknesses, the stable structure is the bcc, so we consider both fcc and bcc structure within a consistent model.
	We determine at what film thickness a structural transition from fcc to bcc can be expected.  
	The results obtained will shed light on the intrinsic properties of the independent Fe film.
	Magnetic films have important uses in various applications, such as spin valves and magnetic tunnelling junctions, of particular interest are magnetic junctions with perpendicular magnetic anisotropy \cite{dieny_perpendicular_2017}.
	The presented study of structural and magnetic properties of Fe ultrathin films is important in the context of preparing more efficient spintronic systems, data storage devices, and magnetic sensors.

\section{Computational details}
	In this study, we examined using density functional theory (DFT) the structural and magnetic properties of iron  ultrathin films with a trigonal-lattice surface, ranging from 3 to 25 atomic monolayers (odd number of atomic monolayers only), see Fig.~\ref{obraz1}. 
	The maximum number of Fe atoms in a supercell is 13 nonequivalent, for a structure composed of 25 atomic monolayers. 
	The calculations were performed with the full-potential local-orbital (FPLO-21) code \cite{koepernik_full-potential_1999-2, opahle_full-potential_1999-1}. 
	FPLO allows for the determination of forces, which is important for the optimization of the structure geometry.
	Our objective was to calculate the magnetic anisotropy energy (MAE) of which the use of full potential method is essential. 
	The exchange-correlation functional in the Perdew-Burke-Ernzerhof (PBE) parametrization was used \cite{perdew_generalized_1996-2}.
	The Brillouin zone was initially sampled using a $20 \times 20 \times 1$ \textit{k}-point mesh to optimize the geometry \cite{monkhorst_special_1976-1}. 
	For all considered systems, the Wyckoff positions were optimized using a scalar-relativistic approach with spin polarization. 
	To find the optimized lattice parameter \textit{a}, the geometry optimization data were interpolated. 
	At equilibrium minima the structures were calculated without forces using a denser \textit{k}-point mesh of $50 \times 50 \times 2$ with the tetrahedron method approach \cite{zaharioudakis_tetrahedron_2004-1}.
	To calculate the magnetic anisotropy energy (MAE) one single step of fully-relativistic calculations was performed for each structure in two quantization directions.
	
\begin{figure*}
	\centering
	\includegraphics[clip, trim =  42 53 21 27, width=0.46\columnwidth]{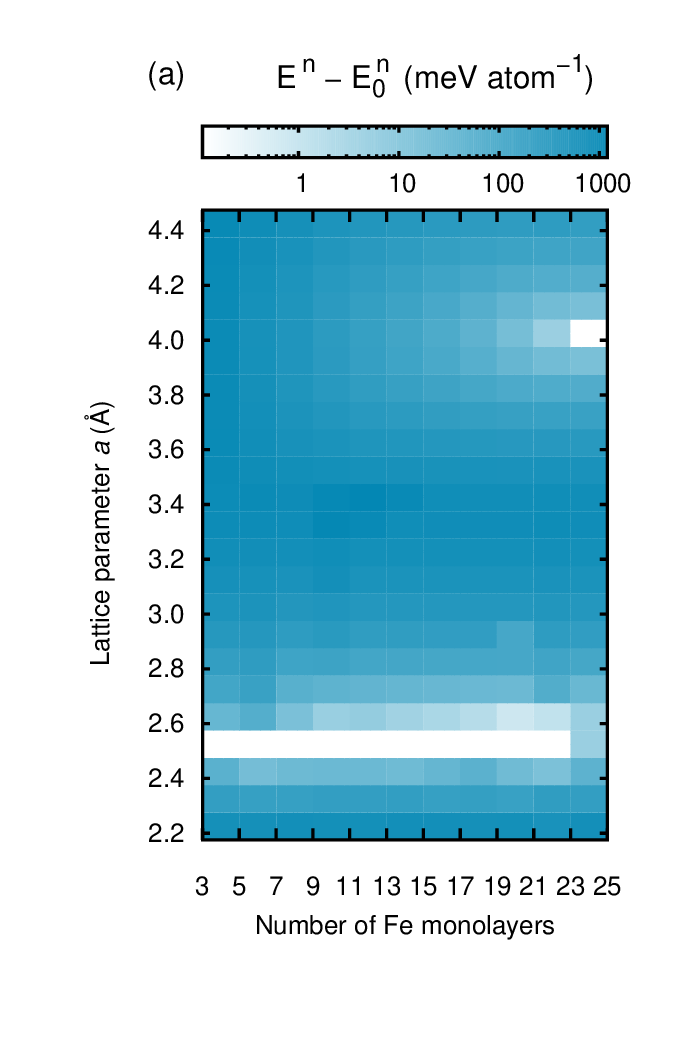}
	\includegraphics[clip, trim =  11 50 31 90, width=0.475\columnwidth]{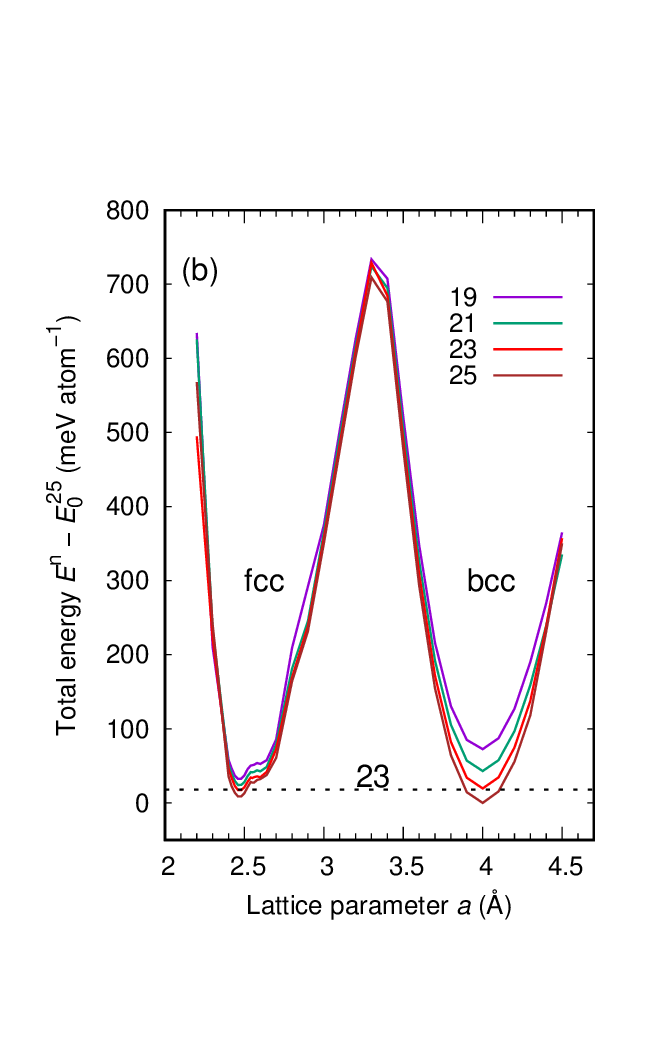}
	\includegraphics[clip, trim =  13 60 25 90 , width=0.385\columnwidth]{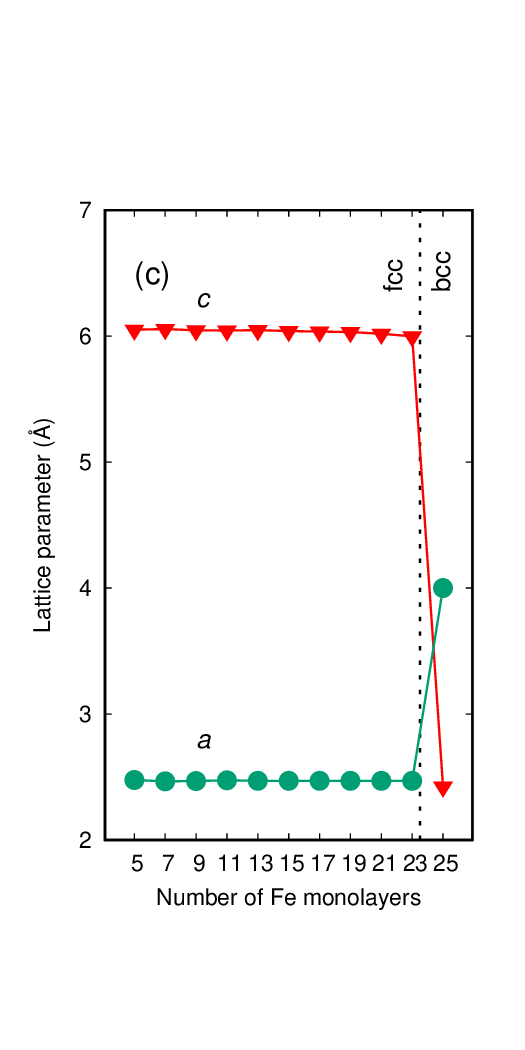}
	\includegraphics[clip, trim =  15 60 25 90 , width=0.38\columnwidth]{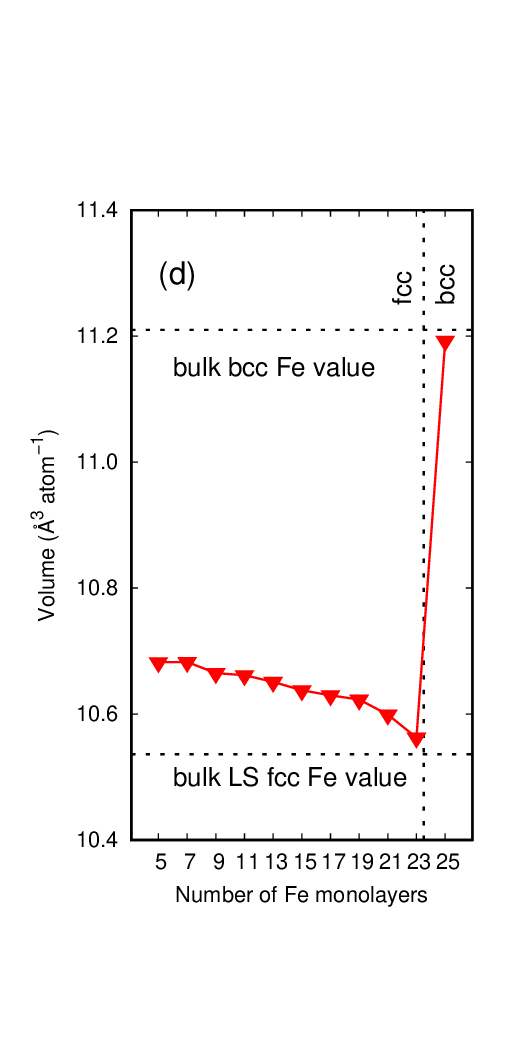}
	\includegraphics[clip, trim =  2 0 2 0 , width=0.295\columnwidth]{LS_HS.eps}
	\caption{\label{obraz2}Structural properties of Fe(111) ultrathin films as a function of number of Fe monolayers and parameter \textit{a} of the trigonal-lattice surface. 
(a) Energy of the system as a function of the number of atomic monolayers and the lattice parameter $a$. Total energies of Fe slabs with \textit{n} monolayers (\textit{E}$^{n}$) shifted to the lowest energy of each slab (\textit{E}$_{0}^{n}$); presented on a logarithmic scale. 
(b) Energy of Fe thin film for several thicknesses around the structural phase transition and as a function of lattice parameter \textit{a} shifted to the lowest energy of the thickest slab (\textit{E}$_{0}^{25}$). 
(c)~Lattice parameters \textit{a} and \textit{c} calculated in the equilibrium state. 
(d) Unit cell volumes compared with values for bulk low-spin fcc Fe and bcc Fe.
(e)~Energy of Fe fcc bulk and thin films in low-spin and high-spin ferromagnetic states.
The DFT calculations were performed with FPLO-21. 
The exchange-correlation functional in the PBE parametrization was used.}
\end{figure*}	
	The MAE values were obtained by comparing the total energy of the system calculated for quantization directions in the film plane and out of film plane.
	The self-consistent scalar-relativistic calculations converged to a tolerance of $10^{-8}$ Ha for the total energy. 
	In each of the systems the structures were fully optimized until the forces on the atoms were less than 0.01 eV\,\AA$^{-1}$.
	To remove the bulk periodicity in the indicated direction and to model the thin film, a vacuum space of at least 15\,\AA $ $ was added.  
	All crystal structures were visualized using the VESTA program\,\cite{momma_vesta_2008-2}.
\section{Results and discussion}

	In our study of the iron films with triangular (111) surfaces, we conducted a thorough analysis of the energy landscape using DFT methods.  
	Our investigation revealed the presence of two distinct minima for the parameter \textit{a} of the trigonal-lattice of the film surface, indicating the existence of two stable configurations for the Fe(111) thin films, see Fig.~\ref{obraz2} (a) and (b).
	These two energy minima occur at about 2.47~\AA{} and 4.00~\AA{}.
	In order to identify which lattice and surface types are characterized by these lattice parameters, we look at the lattice parameters previously obtained for bulk Fe crystals.
	Calculations by Jiang and Carter  \cite{jiang_carbon_2003} for ferromagnetic low-spin bulk Fe fcc gave a lattice parameter value of $a_{bulk.fcc} $ = 3.49 \AA{}. 
	The lattice parameter \textit{a} for fcc Fe(111) films is the length of the cathetus of a right triangle with the hypotenuse of length $a_{bulk.fcc}$, see Fig.~\ref{obraz0}. 
	Therefore, the value of $a_{layer.fcc}$ for an fcc Fe(111) layer taken from the bulk fcc Fe would be $a_{bulk.fcc}/\sqrt{2}$, which is 2.47~\AA{}, based on the above value.
	This value agrees with the lower minimum of the lattice parameter in our results for Fe thin films, see Fig.~\ref{obraz2}.
	
	Furthermore, if we cut a (111) film from bcc Fe, we obtain a structure similar to our fcc (111) film with a lattice defined in the plane by the parameter \textit{a} and an angle of 120 degrees (60 degrees). 
	Experimental data indicate that the lattice parameter of bulk bcc Fe is $a=2.87$~\AA{} and theoretical calculations give $a=2.82$~\AA{} \cite{hsueh_magnetism_2002-1}.  
	Therefore, if we take the calculated value of the lattice parameter $a_{bulk.bcc}$ of bulk bcc Fe to determine the value of the lattice parameter for bcc Fe(111) film, we have $a_{bulk.bcc} \times \sqrt{2}$ = $2.82$~\AA $\times \sqrt{2} \approx 4.00 $~\AA{}, which is exactly the value we see for the second minimum in the energy dependence of the lattice parameter, see Fig.~\ref{obraz2}. 
	It is therefore justified to conclude that the lower minimum corresponds to the fcc lattice and the higher one to the bcc lattice. 
		
	We observe that as the number of atomic monolayers increases above 23, lower global energies begin to be found for systems with a bcc (111) structure. 
	We identify the result  as a structural phase transition from an fcc to a bcc structure. 
	The direction of the transition towards a bcc lattice is intuitively consistent with the bcc structure observed for bulk Fe (large thickness limit). 
	The thickness of the 23 atomic monolayers is 46~\AA{} (4.6 nm) for the Fe fcc (111) layer and only 20.2~\AA{} (2.02 nm) for the Fe bcc (111) layer.
	This implies a significant change in the preferred geometry of the film in the vicinity of the structural transition, which affects the structural properties of ultrathin Fe(111) films with thicknesses in the range of approximately 2 to 5 nm.
	A similar range of critical thickness for the transition from the fcc to the bcc phase in Fe films is observed experimentally, ranging from about 10 to 20 atomic monolayers \cite{kalki_evidence_1993-1, schmailzl_structure_1994-1, hoff_critical_1995-2}.
	
\begin{figure*}
\centering
\includegraphics[clip, trim =  0 40 30 90 , width=0.49\columnwidth]{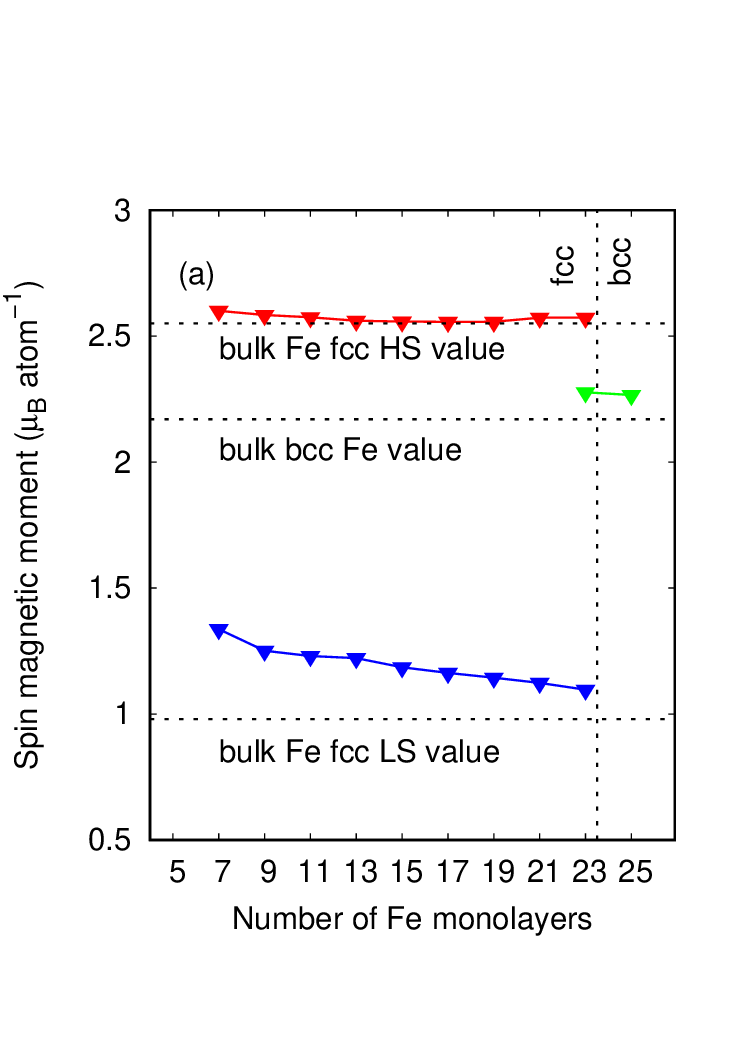}
\includegraphics[clip, trim =  0 40 30 90 , width=0.49\columnwidth]{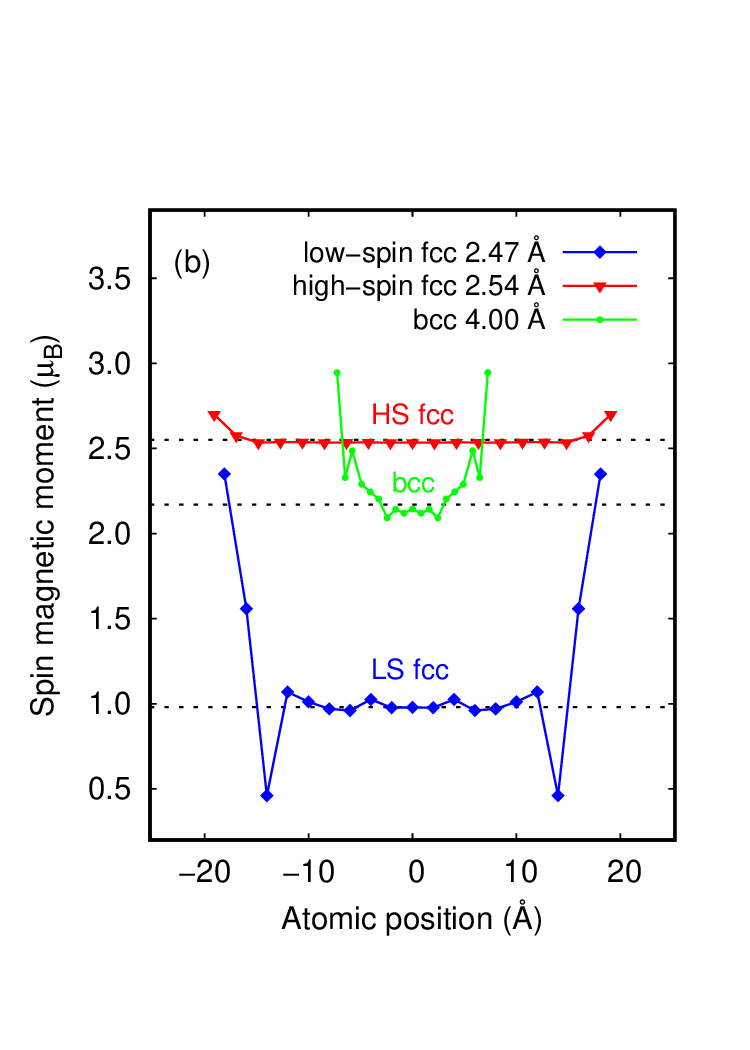}
\includegraphics[clip, trim =  50 40 30 25, width=0.53\columnwidth]{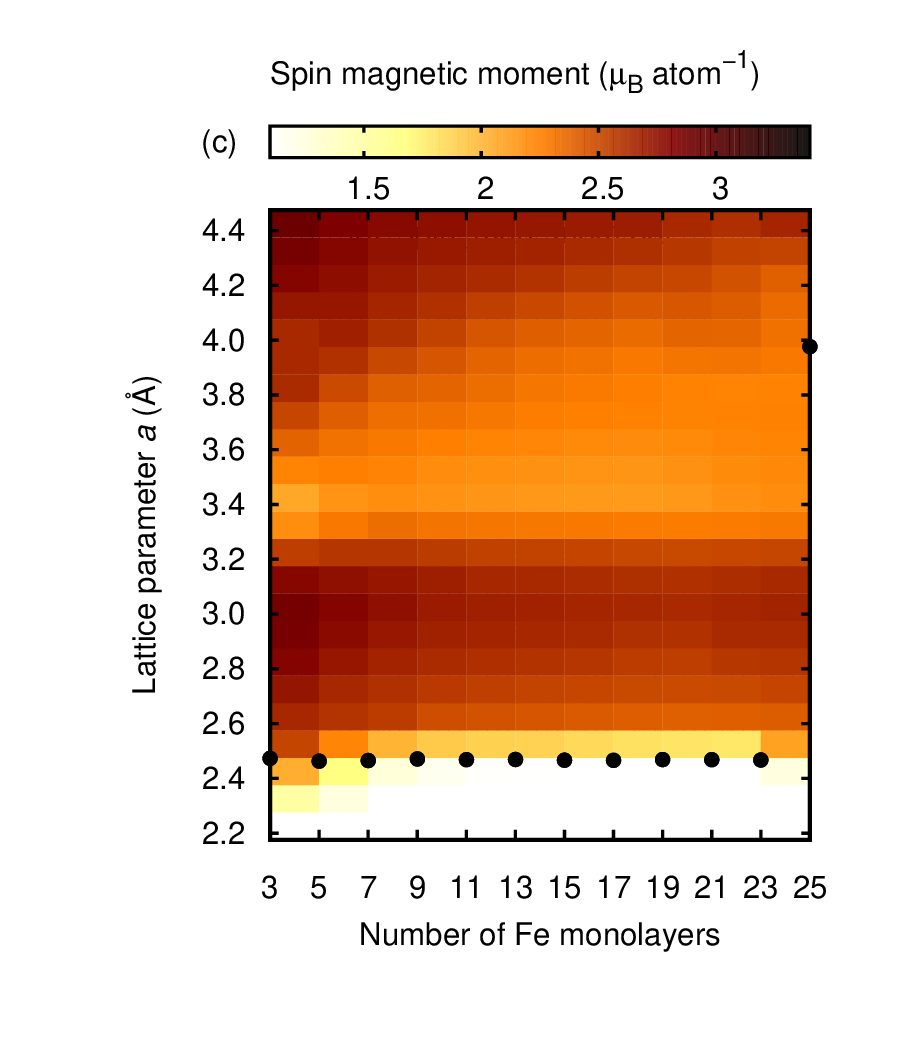}
\includegraphics[clip, trim =  0 40 30 90 , width=0.49\columnwidth]{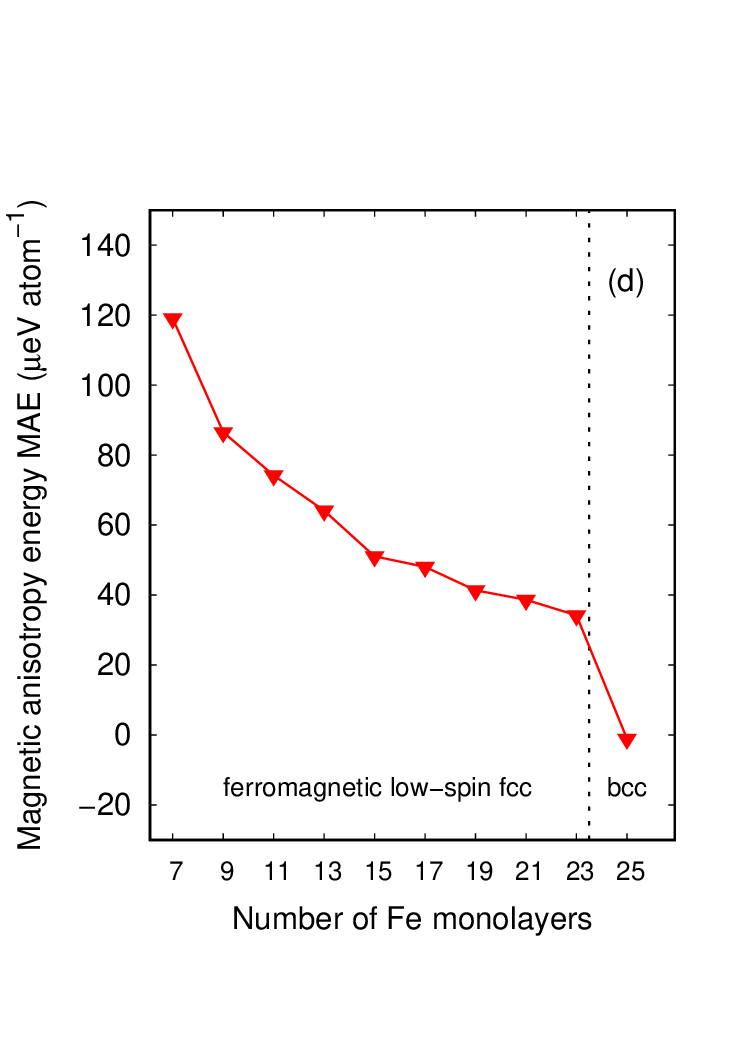}
\caption{\label{obraz3}Magnetic properties of Fe(111) ultrathin films. 
(a) Average spin magnetic moments as a function of the number of Fe monolayers of optimized low-spin and high-spin fcc and bcc films.
(b) Spin magnetic moments at each atomic position of optimized 19-monolayer fcc Fe film in three ferromagnetic magnetostructural states: low-spin fcc ($\tilde{m}_s$ = 1.14 $\mu_B$\,atom$^{-1}$, \textit{a}$_{film}$ = 2.47~\AA{}), high-spin fcc ($\tilde{m}_s$ = 2.56 $\mu_B$\,atom$^{-1}$, \textit{a}$_{film}$ = 2.54~\AA{}), and  bcc ($\tilde{m}_s$ = 2.31 $\mu_B$\,atom$^{-1}$, \textit{a}$_{film}$ = 4.00~\AA{}). 
(c)~Map of total spin magnetic moments as a function of the number of Fe monolayers and the trigonal-lattice surface parameter~\textit{a}, circles on the graph denote calculated equilibrium solutions for bcc and low-spin fcc phases. 
(d) Magnetic anisotropy energy (MAE) as a function of a number of monolayers.
The DFT calculations were performed with FPLO-21. 
The exchange-correlation functional in the PBE parametrization was used.}
\end{figure*}	
	
	The plot of the lattice parameter \textit{a},  see Fig.~\ref{obraz2}(c), shows no significant change with increasing film thickness until the transition point at above 23 monolayers, where it jumps from about 2.47 to 4.00~\AA{}.
	At this point, we define the parameter \textit{c} as the shortest length in the Cartesian direction \textit{z} between periodically occurring atomic positions in the cell.
	The lattice parameter \textit{c} defined in this way includes three atomic monolayers, see Fig.~\ref{obraz1}.  
	With increasing film thickness, the parameter \textit{c} remains  approximately constant around 6.0~\AA{}, while at around 23 Fe monolayers it drops significantly to about 2.4~\AA{}. 
	In fcc Fe deposition experiments on Au(111) substrates, ultrathin fcc Fe films grow with a pseudomorphic lattice parameter \textit{a} of Au(111) equal to 2.88~\AA{} and a lattice parameter \textit{c} equal to 6.27~\AA{} (3$\times$ 2.09~\AA) \cite{stroscio_microscopic_1992-1}.
	Although the values of the lattice parameters obtained experimentally are very close to those determined by us theoretically, we see that it would be possible to match the fcc Fe layer to the substrate even better, if the lattice parameter of the substrate were chosen closer to the 2.47~\AA{} value.
	Changes in the lattice parameters near the structural transition are also reflected in the volume dependence on the number of atomic monolayers in the film.
	In Fig.~\ref{obraz2}(d) we can see that in the vicinity of the phase transition the volume expressed per atom increases stepwise from a value close to bulk fcc Fe to that of bulk bcc Fe.
	Below the structural transition, the volume per atom decreases as the number of monolayers increases.	
	
	Figure~\ref{obraz2}(e) shows the energy minima for the fcc Fe bulk and films.  
	Except for the case of 3-monolayers, we observe two minima separated by a lattice parameter distance of about 0.1~\AA{}. 
	The two structural solutions also differ in their magnetic states, which are referred to in the literature as low-spin and high-spin ferromagnetic states \cite{jiang_carbon_2003}.
	The low-spin (high-spin) state exhibits an average spin magnetic moment of about 1.0 $\mu_B$\,atom$^{-1}$ (2.5 $\mu_B$\,atom$^{-1}$). 
	The result for bulk fcc Fe was previously obtained by Jiang and Carter \cite{jiang_carbon_2003}.
	
	Furthermore, we calculated an average spin magnetic moments of the considered bcc and low-spin and high-spin fcc films, see Fig.~\ref{obraz3}(a). 
	The magnetic moments of the films in these three states resemble the values for their bulk counterparts.
	The elevated average spin magnetic moment values for thinner films result from relatively high surface contributions, and from the proportion of these contributions - increasing with decreasing film thickness. 
	Moreover, we show spin magnetic moments at each atomic position of optimized 19-monolayer fcc Fe films for bcc and low-spin and high-spin fcc states, see Fig.~\ref{obraz3}(b). 
	The plot reveals the distribution of magnetic moments and the difference between the three structures.
	In line with the previously discussed very large variation in the lattice parameter \textit{c} found between fcc and bcc structures, the results presented in the graph for fcc and bcc structures also indicate significant difference in film thickness.
	In all three cases, inside the film we observe values of magnetic moments similar to those for the corresponding phases of bulk Fe, while two to six surface monolayers counted from the surface show significant deviations in the values of the magnetic moments from those in the center of the films.
	The pale dashed lines indicate the values of the magnetic moments for analogous phases of bulk Fe. 
	Moreover, the average values of the spin magnetic moments as a function of the lattice parameter \textit{a} and the number of monolayers are plotted on the map $m_{s}(a,n)$, see Fig.~\ref{obraz3}(c). 
	The map itself reveals the presence of two distinct magnetic regions separated at about $a= 3.3$~\AA{} and  related to two adjacent valleys in the energy landscape, see Fig.~\ref{obraz2}(c).
	Below the lattice parameter \textit{a} of about 2.5~\AA{}, we observe a very large decrease in the value of the average magnetic moment to about 1.0 $\mu_B$\,atom$^{-1}$ and below.
	
	Furthermore, the results show that the positive magnetic anisotropy energy (perpendicular magnetic anisotropy), see Fig.~\ref{obraz3}(d), occurs for the entire range of fcc films (in ferromagnetic low-spin state) and systematically decreases with increasing film thickness.
	At the critical number of monolayers characterizing the change in state preference of the film from low-spin fcc to bcc (about 23), the MAE drops to a value close to zero, which is characteristic of bulk bcc Fe.
	
\section{Summary and conclusions}
	This work investigates the structural phase transition and magnetic properties in ultrathin Fe films with a trigonal-lattice surface, focusing in particular on Fe(111) fcc and bcc thin films.
	The energy landscape of these structures was found to exhibit energy minima for two distinct values of lattice parameters in the plane of the films. 
	The minimum at around 2.47~\AA{} is considered as ferromagnetic low-spin fcc  Fe(111) and that at around 4.0~\AA{} as ferromagnetic bcc Fe(111).
	We have shown that in ultrathin Fe films with a trigonal-lattice surface, a structural phase transition from fcc to bcc Fe occurs at a thickness of about 23 atomic monolayers. 
	At the structural transition we observe step changes in the values of the lattice parameters and consequently in the film volume.
	The results are in agreement with the experimentally observed fcc-bcc structural transition in ultrathin Fe films.
	However, our calculations show that the more stable fcc structure observed for the lower Fe(111) thicknesses is not the result of stabilization by the substrate, but the ground state of the free-standing Fe film.
	In the fcc structure, we observe ferromagnetic low-spin and high-spin states, with an energy minimum for the former.
	The average magnetic moment in low-spin fcc Fe films converges with increasing number of monolayers toward the value determined for bulk low-spin fcc Fe.
	The values of the magnetic anisotropy energy are positive (perpendicular magnetic anisotropy) for the entire thickness range of the low-spin fcc phase and systematically decrease with increasing film thickness.

\section*{Acknowledgments}
	We acknowledge the financial support of the National Science Center Poland under the decision DEC-2018/30/E/ST3/00267 (SONATA-BIS 8).
	Part of the computations were performed on resources provided by the Poznan Supercomputing and Networking Center (PSNC).
	We thank Justyn Snarski-Adamski and Igor Di Marco for valuable comments and Paweł Leśniak and Daniel Depcik for compiling the scientific software and administration of the computing cluster at the Institute of Molecular Physics, Polish Academy of Sciences.

\bibliography{test4.bib}    

\end{sloppypar}
\end{document}